\documentclass[runningheads]{llncs}
\usepackage[T1]{fontenc}
\usepackage{graphicx}
\usepackage{booktabs}
\begin{document}
\title{Profiling User Vulnerability to Phishing Through Psychological and Behavioral Factors}

 \author{
    Valeria Formisano\inst{1} \and
    Danilo Gentile\inst{1}\inst{2} \and
    Gennaro Esposito Mocerino\inst{1} \and
    Michela Ponticorvo\inst{1} \and
    Luigi Gallo\inst{1}\inst{3} \and
    Alessio Botta\inst{1} \and
    Davide Marocco\inst{1}
}
 \institute{
     Department of Electrical Engineering and Information Technology,
     University of Naples Federico II, Naples, Italy\\
     \email{\{gennaro.espositomocerino, alessio.botta, giorgio.ventre, michela.ponticorvo, davide.marocco\}@unina.it},
     \and Cyber Security Fibercop S.p.A., Naples, Italy
     \email {danilo.gentile@fibercop.com}
     \and Cyber Security TIM S.p.A., Turin, Italy
     \email {luigi1.gallo@telecomitalia.it}
 }
 \authorrunning{Formisano et al.}

\titlerunning{Profiling User Vulnerability to Phishing Through Psychological and Behavioral Factors}
\maketitle              % typeset the header of the contribution
\begin{abstract}
Phishing remains one of the most pervasive cybersecurity threats, shifting the focus from technological vulnerabilities to human cognitive and psychological factors. In coherence with the trend of studies on phishing to increasingly focus on human aspects and vulnerable users profiling, this study investigates the multidimensional nature of user susceptibility by analyzing data from the Spamley dataset, involving 1,086 participants evaluated through a realistic phishing detection task. Using Exploratory Factor Analysis (EFA), five latent constructs were identified, named: Seniority, Expertise, Creativity, Stability, and Vulnerability. Behavioral findings, validating self-reported impulsivity through its negative correlation with response times, demonstrate that faster decision-making significantly distinguishes vulnerable users from resilient ones. A K-Means clustering procedure, driven by the dimensions of Seniority (F1) and Creativity (F3), revealed two distinct user profiles: the Aware User and the High-Risk User. The results demonstrate that technical knowledge alone is insufficient to guarantee resilience; rather, the interaction between operational maturity, decision-making speed, and cognitive approach determines effectiveness. The findings suggest that the majority of users fall into the High-Risk category, characterized by hasty evaluation processes and lower critical analysis. These results underline the urgent need to move beyond "one-size-fits-all" training toward personalized, adaptive cybersecurity programs that actively address cognitive biases and behavioral tendencies.
\keywords{Cybersecurity \and Phishing email \and Cognitive Vulnerability \and Security Awareness \and Human Factor}
\end{abstract}

\section{Introduction and Motivation}
\label{sectionIntroduction}
With the continuous expansion of digital services and communication networks, sharing personal and financial information online has become an everyday practice, exposing users to increasingly sophisticated threats. In this digital landscape, phishing stands out as one of the most widespread and effective forms of cyber crime, as it allows attackers to bypass purely technological defenses by targeting the end point of every technological system directly: the user \cite{alkhalil2021phishing}. Traditional cybersecurity measures, however advanced, are often insufficient unless integrated with a deep understanding of human behavior, which constitutes the true target of these threats \cite{moustafa2021behavior}.
The success of phishing attacks, understood as complex social engineering tactics, fundamentally depends on the interaction of two distinct dimensions: on the one hand, the characteristics of the message, and on the other, the characteristics of the person receiving it. 

With respect to the first aspect, the effectiveness of the attack is determined by a set of structural and psychological elements carefully designed to deceive the victim. At the structural level, the characteristics of the email play a crucial role in the perception of authenticity: cybercriminals use spoofing techniques to forge the sender’s address, faithfully replicate the aesthetics, logos, and formatting of legitimate organizations (such as banks or government agencies), and disguise malicious URLs or infected attachments by presenting them as routine documents, such as invoices or receipts\cite{Alsharnouby2023,Abdelhamid2021}.
Psychologically, these communications are structured around specific principles of persuasion developed in social psychology by Robert Cialdini \cite{Cialdini1993}, used to circumvent the user’s critical thinking. For example, attackers frequently exploit authority, impersonating corporate executives, IT administrators, or law enforcement to intimidate the victim and coerce immediate compliance. Similarly, they use the principle of reciprocity by offering fake rewards or unexpected refunds, or they leverage consistency, first requesting small, seemingly harmless actions and then pushing the user to hand over sensitive credentials \cite{stylianou2025suspicious}. It is worth noting that many phishing attacks make use of urgency (such as the threat of imminent account suspension within 24 hours) to induce tension and force hasty decisions, yet quantitative studies show that the use of levers such as authority and consistency \cite{stylianou2025suspicious} or appreciation (also known as liking) \cite{lawall2025persuasion} are actually the strongest predictors of compromising the user’s defenses.

However, vulnerability to such attacks is not limited to the effectiveness of the deceptive strategies employed in the message, but also on the characteristics of the human that reads the email. This study therefore decided to focus its initial analysis on this still largely unexplored phenomenon: the human psychological factor. Specifically, our initial goal was to identify, within a publicly available dataset on phishing susceptibility, the profiles of users who are potentially most vulnerable to phishing attacks. To this end, we analyze data from the \textit{Spamley} dataset \cite{spamleywebapp}, a publicly available dataset on users' ability to recognize phishing emails. Susceptibility to fraud stems from a complex interaction between the external attack and the individual’s intrinsic decision-making, cognitive, and physiological processes. Understanding the human being behind the screen is essential for developing effective defense models.

\subsection{The State of the Art}
\label{SectionRelatedWorks}
An analysis of the existing literature reveals that research has progressively explored the human factor by cataloging a wide range of predictive variables \cite{tornblad2021characteristics,kavvadias2025review}.
A first fundamental line of research has focused on cognitive and decision-making processes. It has been widely demonstrated that information processing modes play a key role. In fact, the likelihood of falling victim to fraud increases significantly when users rely on heuristic processing referred to as “System 1”  \cite{kahneman2011thinking} and described as intuitive, emotional, and fast, rather than on a systematic and rational analysis of the message (“System 2”)\cite{xu2025fraud,frauenstein2020phishing}. Added to these dynamics is the pervasive effect of habit: the automatic and routine use of email can inhibit the activation of suspicion, lowering the cognitive barriers necessary to detect anomalies and deception \cite{vishwanath2018scam}. Conversely, expertise in the subject allows for the development of effective defensive heuristics, such as the ability to interpret linguistic anomalies or instinctively recognize unrealistic offers in everyday life \cite{aldaraani2023ksa}.
A second crucial strand concerns psychological and personality traits. Much research draws on the Big Five psychometric model, confirming that fundamental personality traits directly influence how messages are processed. For example, high levels of extroversion, openness, or neuroticism tend to correlate with greater susceptibility to certain phishing scenarios, such as those promising rewards or inducing emotional stress \cite{gordon2024bigfive,lawson2020signal,halevi2013pilot,frauenstein2020phishing}. Furthermore, intrinsic signal detection abilities and the level of suspicion (understood both as a dispositional traits inherent in the individual, and as a transient psychological state induced by the context) profoundly alter judgments of truthfulness and accuracy in detecting deception \cite{takiguchi2025suspicion}. While some approaches make use of unsupervised learning models to uncover vulnerable users profiles \cite{wafik2025}, others focus on user behavior while performing specific tasks related to phishing content analysis \cite{pietrantonio2024,mocerino2024}.
Finally, the literature has explored the impact of demographic, physiological, and contextual variables. Vulnerability trajectories change drastically with age. Aging, especially when associated with a decline in executive functions or specific genetic profiles, tends to severely exacerbate susceptibility to phishing \cite{pehlivanoglu2024phishing}. However, compensatory factors have also been identified: high interoceptive accuracy (the ability to perceive one’s body’s internal signals, such as heart rate) can improve deception detection skills in older adults \cite{heemskerk2024interoceptive}. At the opposite end of the age spectrum, the psychological impact also hits young adults hard, among whom direct experience with cyber attacks generates anxieties and concerns that have proven to be cross-cultural and universal \cite{aldaraani2023ksa}. 
% Furthermore, as communication shifts to alternative platforms, the scope of risk has evolved: on instant messaging apps, relational dynamics, prior trust in the sender, and peer pressure increase the likelihood of interacting with malicious links \cite{ahmad2022phishing}, while on social media, specific variables such as professional role and age directly determine exposure to threats \cite{Salamah2023}. 
Numerous researches demonstrated the association between personality traits and phishing vulnerabilities, beyond technical reasons \cite{moustafa2021behavior,tornblad2021characteristics}. For example, impulsive behavior were shown to be associated with higher chances of deception \cite{hadlington2017,welk2015,kavvadias2025}. Our study fits into this line of research, aiming to further refine user profiling. The primary goal is to identify those specific characteristics that can describe and be associated to a given individual’s vulnerability to phishing; in other words, we aim to obtain a reliable estimate of an individual’s risk exposure based on the traits that describe them.

\subsection{Research Gap}
%inserire altri lavori di profilazione e le loro mancanze
In light of the complex cognitive, psychological, demographic, and contextual dynamics involved in phishing detection, this study aims to conduct an in-depth analysis of human factors, with the primary objective of delineating the typical profile of an individual vulnerable to phishing. While the concept of user profiling has increasingly gained traction in cybersecurity research to understand victimization patterns and tailor interventions \cite{ge2021personal,kavvadias2025review,pehlivanoglu2024phishing}, there remains a critical need for an integrated and multidimensional approach to accurately identify profiles of vulnerable and resilient individuals with regard to phishing. A key strength of our approach lies in the use of a large public dataset. Unlike many previous studies in this field, which are often constrained by small samples that can limit the generalizability of results, this vast amount of data allows us to conduct more rigorous and scalable user profiling analyses. This ensures statistically sound and robust results, providing a solid empirical foundation for the construction of future predictive models. Therefore, this research aims to identify which specific psychological variables (e.g., personality traits, risk perception, cognitive styles) and behavioral variables (e.g., communication management habits, impulsivity, response speed) are intrinsically associated with higher chances of victimization. Ultimately, the study seeks to answer the following research question:\\

\noindent \textit{\textbf{RQ}} - What human factors and combinations of psycho-behavioral variables best describe a user’s vulnerability to phishing?\\

\noindent By filling this gap in the literature, the study aims to provide a solid and up-to-date empirical basis for the development of defensive countermeasures, corporate policies, and security training programs. By leveraging evidence-based user profiling methodologies, these countermeasures can move beyond a generic “one-size-fits-all” approach to become increasingly personalized, targeted, and effective.

\section{Methodology}
\label{sectionMethodology}
To pursue the study’s objectives and ensure robust statistical validity, a heterogeneous sample of 1,086 participants was recruited, spanning various socio-professional backgrounds, creating a highly multidimensional dataset. Each subject was administered a structured questionnaire specifically designed to map and investigate a broad spectrum of sociodemographic, psychological, educational, and behavioral variables. As a result of the answers provided by participants to the questionnaire and the phishing test, the dataset analyzed in this study was created. On one hand, the dataset contains the users’ responses to the questionnaire items. It also includes data on users’ ability to distinguish phishing emails from legitimate communications; details of the classification task performed by users are provided in Section \ref{Test}. Finally, the dataset incorporates all the features of the emails presented during the test, ranging from visual features to structural characteristics. Gallo et al. \cite{gallo2024} provide a thorough explanation of the questionnaire items composing the dataset used in this study, as well as a detailed description of the recruiting process and the general composition of the \textit{Spamley dataset}, for both the user and email components. For the purposes of this research work, the dataset was enriched with data derived from the classification task: for each user, the median response time and recall (i.e., the percentage of correctly classified phishing emails) were calculated.

\subsection{The phishing test}
\label{Test}
\begin{figure}[t]
    \centering
    \includegraphics[alt={A possible email presented to the user during the test.},width=0.80\textwidth]{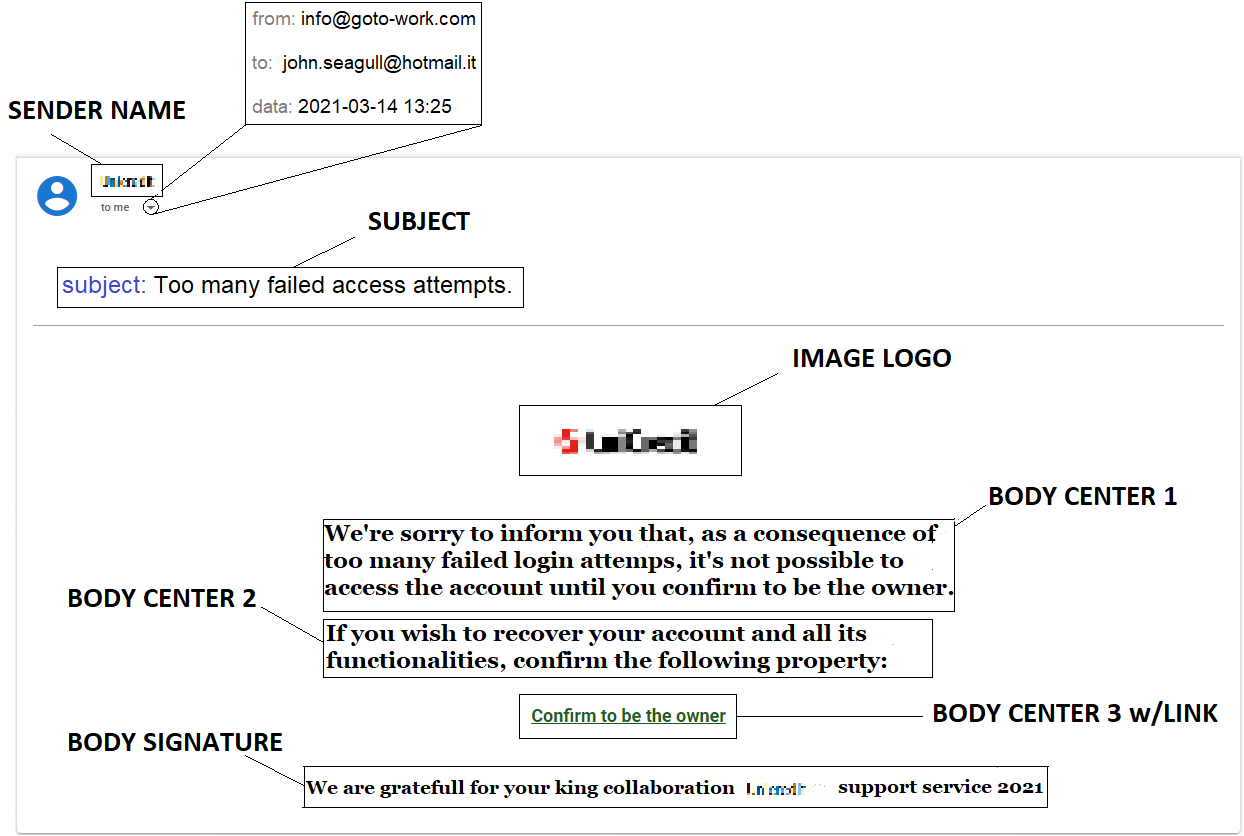}
    \caption{A possible email presented to the user during the test. Each email is composed of several fields, which are characterized by different special features in the dataset.}
    \label{fig:phishing_screen}
\end{figure}

To collect behavioral data and evaluate the sample’s performance, the study employed a publicly available web app called Spamley \cite{spamleywebapp}, a system specifically designed to collect and analyze data on user behavior while reading emails. Through this web application, developed to test individuals’ awareness of the phishing threat, participants underwent a practical simulation in which they were asked to examine a series of ten email messages. For each email, participants had to determine whether it was a legitimate communication or a fraudulent attempt. To ensure the accuracy of the collected data and mitigate the tendency of participants to alter their behavior —generally becoming more suspicious when they know they are exposed to phishing attempts— the system implemented specific design strategies. First, the simulation’s graphical interface was structured to faithfully replicate a real email client, incorporating numerous visual elements similar to those of a very popular web email client. The web application also supports both desktop and mobile interfaces, allowing participants to take the test from their own computer or smartphone, depending on their usual communication-reading scenario. These measures aim to place users in an operating environment that is fully familiar to them. Finally, to maintain a high degree of realism, the messages presented were carefully crafted to be as similar as possible to emails received in real-world contexts, and an introductory page clarified to participants that the receipt of messages (legitimate or malicious) was entirely random, exactly replicating the dynamics of a real-world scenario. Figure \ref{fig:phishing_screen} shows the details of a typical email that users receive during the test.

\subsection{Data Preprocessing}
An initial dataset was generated from the raw data collected via the Spamley web platform and the integrated pre-test questionnaire. Subsequently, a rigorous data cleaning and pre-processing procedure was implemented to isolate only the records deemed valid for statistical analysis. The validity and inclusion criteria required the complete provision of both the information in the questionnaire and the classification activity on the test, resulting in the automatic exclusion of incomplete sessions. Furthermore, to ensure data reliability and mitigate the risk of bias associated with random or inattentive completion, an attention control item was introduced. This item explicitly required the subject to enter the value “3” to demonstrate that they maintained an adequate level of attention during the administration. Users were selected on the basis of the provided attention check value and the time to complete the questionnaire. By analyzing the timestamps of the login on the web application and the start of the test for each user, it was possible to calculate how much time the user spent on the initial web page and on the questionnaire; for users with the attention check value "3" the median time spent was 4 minutes, so we filtered out from the dataset all users with attention check "3" that took less than 2 minutes, as we have observed it's a largely insufficient time to complete the questionnaire while paying attention, and included all users with attention check different from 3 that took 4 or more minutes, which is above average for the "valid" group. Finally, 10 users which received less than 3 phishing emails were excluded, to improve individual statistical relevance. At the conclusion of this thorough filtering phase, the final dataset was consolidated, consisting of a sample of 1,086 eligible participants, on which all subsequent analyses were conducted.

\subsection{Data Analysis Tools}
Data processing and analysis were conducted using the statistical software Jamovi and the Python programming language. On the variables under study were performed an extraction of statistical properties of the questionnaire answers to derive some descriptive analysis, a bi-variate analysis to observe any correlation between user characteristics and phishing resilience, an Exploratory Factor Analysis (EFA) and finally a clustering algorithm to derive user profiles. The descriptive analyses outlines the distribution of demographic, psychological, and behavioral variables. Through bi-variate analysis, correlations between these variables and participants’ performance were explored, with the aim of identifying constructs significantly associated with the recall rate. Furthermore, to corroborate the validity of the self-reported measure of impulsivity, its correlation with a related behavioral index —response times— was analyzed. The underlying hypothesis posited that higher self-reported impulsivity would correspond to significantly shorter response times. In line with this premise and with current evidence in the literature \cite{moustafa2021behavior}, potential significant differences in response times between the victim group and the resilient group were explored. Then a factors extraction was performed using the Minimum Residual method with \textit{Oblimin oblique rotation}, determining the number of factors to retain via Parallel Analysis. Finally, the sample was segmented using the K-Means clustering algorithm. To identify the optimal number of clusters (profiles), models with a number of clusters ranging from 2 to 6 were evaluated. The final $k$ was chosen by selecting the configuration that yielded the highest Silhouette coefficient.

% Data processing and analysis were conducted using the statistical software Jamovi and the Python programming language. Exploratory Factor Analysis (EFA) was performed on the variables under study using Jamovi. Factor extraction was performed using the Minimum Residual method with Oblimin oblique rotation, determining the number of factors to retain via Parallel Analysis. Python was used for descriptive statistics, bi-variate analysis, and sample profiling. The descriptive analyses focused on the distribution of demographic, psychological, and behavioral variables. Through bi-variate analysis, correlations between these variables and participants’ performance were explored, with the aim of identifying constructs significantly associated with the recall rate. Furthermore, to corroborate the validity of the self-report measure of impulsivity, its correlation with a related behavioral index—response times—was analyzed. The underlying hypothesis posited that higher self-reported impulsivity would correspond to significantly shorter response times. In line with this premise and with current evidence in the literature [15], potential significant differences in response times between the victim group and the resilient group were explored. Finally, the sample was segmented using the K-Means clustering algorithm. The identification of the optimal number of clusters (profiles) was guided by the calculation of the Silhouette coefficient.

\section{Analysis of Results}
\label{sectionResults}
This section will provide an outline of the analysis conducted on the object dataset, highlighting its characteristics and the key findings. Most participants are Italian, which may limit generalizability across different cultures; still, the sample spans a wide range of professional roles, fields of study, ages, and personality traits, supporting ecological validity.
\subsection{Descriptive Statistics and Bivariate Analysis} The preliminary analysis allowed us to calculate the mean per-user recall, defined as the average proportion of correctly identified phishing emails across users, which stood at 0.716. The bivariate analysis revealed several significant associations between the ability to correctly identify phishing emails (recall) and the sociodemographic, psychological, and behavioral variables investigated. Specifically, Table \ref{tab:correlations} lists all the relevant correlations between victimization and variables.
\begin{table*}[t]
\centering
\scriptsize
\renewcommand{\arraystretch}{1.1}
\begin{tabular}{lrrrrrrl}
\toprule
\textbf{Variable} & \textbf{Mean} & \textbf{Std. Dev.} & \textbf{Min} & \textbf{Max} & \textbf{Rho} & \textbf{P-value} & \textbf{Significance} \\
\midrule
Age                              & 31.00 & 13.17 & --    & --    & 0.251  & $<$0.001 & Sig. \\
Years of Job Experience          & 7.78  & 10.65 & --    & --    & 0.241  & $<$0.001 & Sig. \\
Computer Science Knowledge       & 3.10  & 1.32  & 0.00  & 5.00  & 0.185  & $<$0.001 & Sig. \\
Education Level                  & 2.10  & 1.07  & 1.00  & 4.00  & 0.163  & $<$0.001 & Sig. \\
Work Hours Prior to Test         & 3.00  & 2.94  & 0.00  & 10.00 & 0.161  & $<$0.001 & Sig. \\
Gender                           & --    & --    & --    & --    & 0.132  & $<$0.001 & Sig. \\
Self-confidence                  & 3.49  & 1.21  & 0.00  & 5.00  & 0.107  & $<$0.001 & Sig. \\
Authority (persuasion)           & 3.26  & 1.11  & 0.00  & 5.00  & -0.084 & 0.006 & Sig. \\
Agreeableness                    & 3.62  & 1.12  & 0.00  & 5.00  & -0.079 & 0.009 & Sig. \\
Prior Phishing Attack Experience & 0.16  & 0.36  & 0.00  & 1.00  & -0.072 & 0.017 & Sig. \\
Consistency (persuasion)         & 2.91  & 1.28  & 0.00  & 5.00  & 0.069  & 0.023 & Sig. \\
Anti-phishing Course             & 0.07  & 0.26  & 0.00  & 1.00  & 0.066  & 0.030 & Sig. \\
Emotional Stability              & 3.00  & 1.24  & 0.00  & 5.00  & 0.061  & 0.043 & Sig. \\
\midrule
Openness                         & 3.67  & 1.12  & 0.00  & 5.00  & -0.059 & 0.052 & n.s. \\
Risk Perception                  & 3.47  & 1.06  & 0.00  & 5.00  & 0.058  & 0.056 & n.s. \\
Conscientiousness                & 3.34  & 1.12  & 0.00  & 5.00  & 0.051  & 0.094 & n.s. \\
Curiosity                        & 2.67  & 1.47  & 0.00  & 5.00  & -0.045 & 0.138 & n.s. \\
Scarcity                         & 2.51  & 1.21  & 0.00  & 5.00  & -0.044 & 0.144 & n.s. \\
Impulsivity                      & 2.12  & 1.25  & 0.00  & 5.00  & -0.044 & 0.151 & n.s. \\
Time on Internet                 & 5.84  & 2.46  & 1.00  & 10.00 & 0.035  & 0.242 & n.s. \\
Extroversion                     & 2.61  & 1.36  & 0.00  & 5.00  & -0.018 & 0.560 & n.s. \\
Social Proof                     & 2.27  & 1.17  & 0.00  & 5.00  & -0.015 & 0.624 & n.s. \\
Risk Propensity                  & 1.89  & 1.28  & 0.00  & 5.00  & 0.007  & 0.810 & n.s. \\
Privacy Concerns                 & 3.51  & 1.29  & 0.00  & 5.00  & -0.002 & 0.943 & n.s. \\
Gratitude                        & 3.42  & 1.08  & 0.00  & 5.00  & -0.002 & 0.960 & n.s. \\
\bottomrule
\end{tabular}
\caption{Spearman correlations with descriptive statistics. Many questionnaire items correlate with the ability to recognize phishing emails, notably age, job experience, computer skill and self confidence in spotting malicious emails.}
\label{tab:correlations}
\end{table*}

% \subsubsection{Key takeaway - Recall Correlations} Many questionnaire items correlate with the ability to recognize phishing emails, notably age, job experience, computer skill and self confidence in spotting malicious emails.

\subsection{Validation of Impulsivity and Response Times} To corroborate the validity of the impulsivity measurement, its relationship with task response times —considered a direct behavioral manifestation of the construct— was analyzed. As expected, a significant negative correlation (r = -0.177, p < 0.001) emerged between self-reported impulsivity and reaction times. Finally, to investigate possible differences in reaction times between the most vulnerable individuals (“victims”) and those best able to recognize the threat (“resilient individuals”), an independent samples t-test was conducted. The analysis revealed a statistically significant difference between the two groups (t = -4.198, p < 0.001), confirming that the victim group and the resilient group exhibit dissimilar response time patterns.
\subsection{Exploratory Factor Analysis (EFA)} In order to explore the latent structures within the set of collected variables, an Exploratory Factor Analysis was conducted. Factor extraction was performed using the Minimum Residual method, combined with Oblimin oblique rotation to facilitate the interpretation of factor loadings. The analysis led to the identification of five main latent factors (defined by authors as Seniority, Expertise, Creativity, Stability, and Vulnerability), whose loadings describe the aggregation of the various sociodemographic, psychological, and behavioral variables.
\begin{figure}[t]
    \centering
    \includegraphics[alt={Heatmap of questionnaire items weights relative to the 5 identified factors.},width=0.85\textwidth]{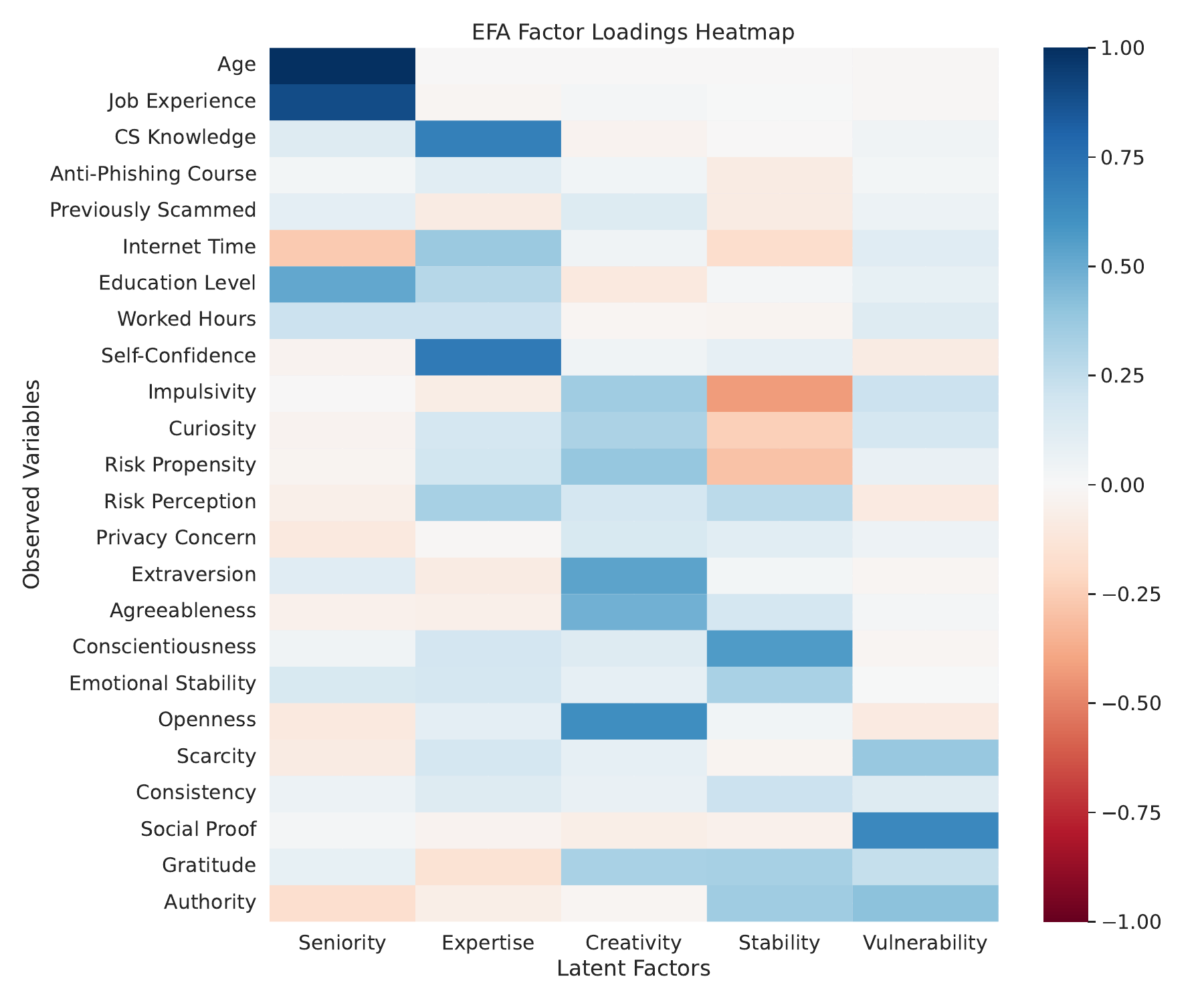}
    \caption{\textit{Heatmap of questionnaire items weights relative to the 5 identified factors. Questionnaire items aggregate best in 5 factors, with seniority, expertise and creativity having the strongest loadings.}}
    \label{fig:STEM}
\end{figure}

\subsection{User Profiling via K-Means Clustering} 
\begin{figure}[h]
    \centering
    \includegraphics[alt={Factors distribution of the two identified profiles.},width=0.85\textwidth]{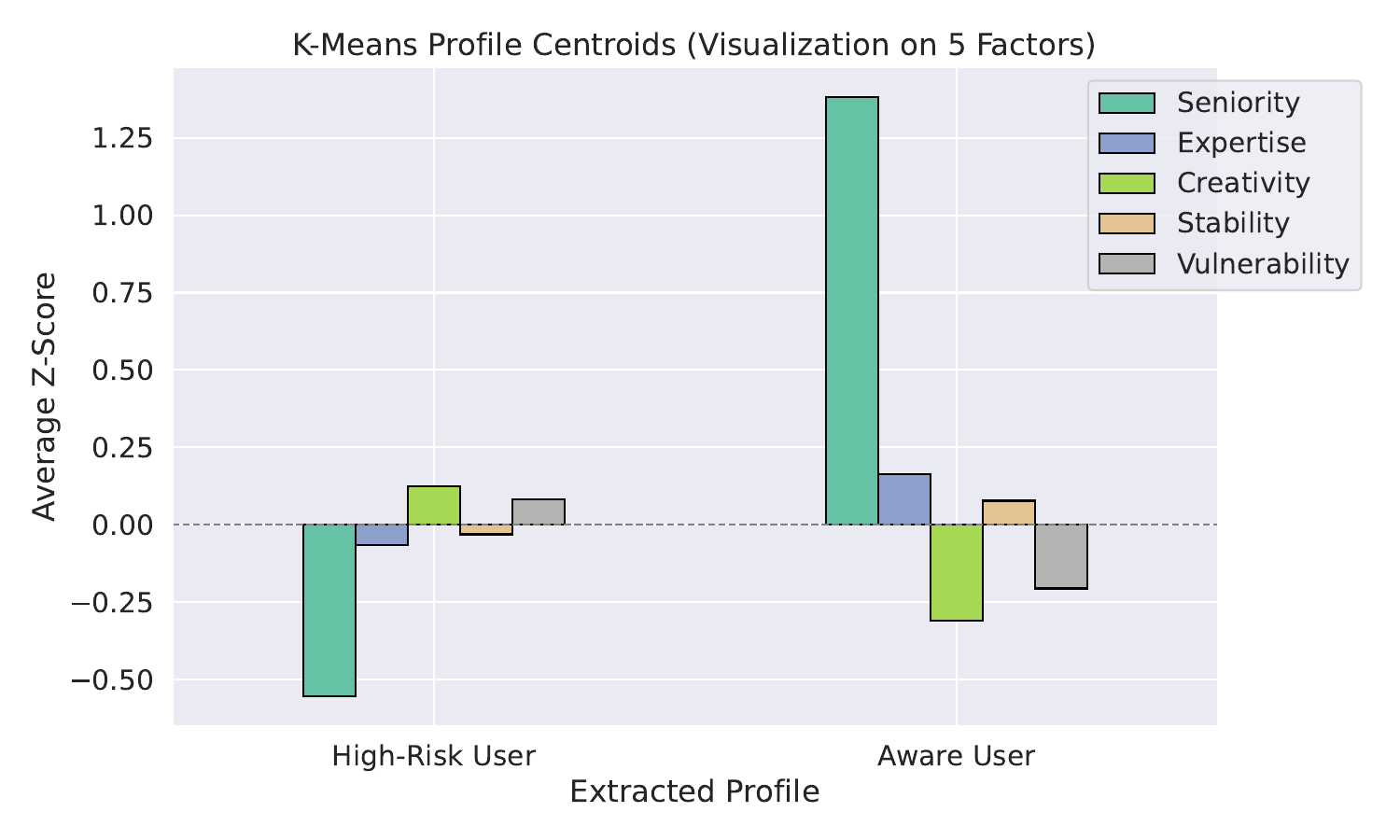}
    \caption{Factors distribution of the two identified profiles. The factors that best model the profiles of resilient users and users at risk are the ones related to seniority and openness.}
    \label{fig:profiles}
\end{figure}

To identify homogeneous clusters within the sample, a profiling analysis was conducted using the K-Means unsupervised learning algorithm. The clustering procedure was performed using two specific latent variables derived from the Exploratory Factor Analysis: Seniority (F1) and Creativity (F3). This dimensional selection was driven by the optimization of the Silhouette Coefficient, which yielded a value of 0.479, ensuring a robust and cohesive cluster structure. The analysis determined an optimal solution consisting of two distinct profiles (k = 2). While the K-Means algorithm utilized exclusively the Seniority and Creativity dimensions for partitioning, the remaining three factors —Expertise (F2), Stability (F4), and Vulnerability (F5)— were integrated post-clustering as descriptive variables. This allowed for a more comprehensive psychometric and behavioral characterization of the emerged groups. The two profiles shown in Figure \ref{fig:profiles} are detailed as follows: \textbf{High-Risk User} (71.4\%, N = 775): Representing the majority of the sample, this group exhibits the highest vulnerability to threats. It is characterized by the lowest baseline recall rate (68.8\%) and faster, potentially more impulsive, average reaction times (21.4s). Psychometrically, this profile shows a negative score on Seniority (-0.55), indicating a demographic tending toward younger age and fewer years of job experience. This is coupled with a slightly positive score on Creativity (0.12), which reflects higher levels of openness, extroversion, and risk propensity. \textbf{Aware User} (28.6\%, N = 311): This group represents the more resilient cluster. These users achieved a significantly higher recall rate (78.3\%), supported by a more analytical approach as reflected in longer average reaction times (23.6s). This profile is strongly defined by a high positive score on Seniority (1.38), pointing to older individuals with greater professional experience and education levels. Furthermore, their negative score on Creativity (-0.31) indicates lower impulsivity and a reduced propensity for taking risks. Although not employed for the initial algorithmic partitioning, the supplementary factors further differentiate the behavior of the two clusters and provide crucial context regarding their cognitive frameworks: the Aware User demonstrates positive scores on Expertise (0.16) and Stability (0.08), reflecting higher self-confidence, better computer science knowledge, and strong conscientiousness. Alongside this, a negative orientation on Vulnerability (-0.21) highlights a robust resistance to common social engineering tactics, such as scarcity and social proof. Conversely, the High-Risk User presents slightly negative scores on Expertise (-0.07) and Stability (-0.03), suggesting lower baseline technical knowledge and emotional stability. Critically, they maintain a positive score on Cognitive Vulnerability (0.08), indicating a heightened susceptibility to psychological triggers like authority and social proof. Overall, the clustering results indicate that specific configurations of Seniority and Creativity correlate directly with increased caution (longer RTs) and better phishing email detection (higher recall). The targeted factor selection (Silhouette = 0.479) provided a highly interpretable and statistically sound framework for delineating user vulnerability, clearly linking demographic and personality traits to actual cybersecurity performance.

% \subsubsection{Key takeaway - User profiling} The factors that best model the profiles of resilient users and users at risk are the ones related to seniority and openness.

\section{Discussion, Practical Implications and Future Works}
\label{sectionDiscussion}
A preliminary exploration of correlations reveals that the ability to recognize phishing threats is supported by a combination of demographic and professional maturity (age, years of work experience) and personality traits (Openness, extroversion, agreeableness, risk propensity). It is particularly interesting to note the negative correlation between susceptibility to phishing and personality traits linked to social conformity, such as authority and agreeableness. This suggests that individuals more inclined to please or obey authority are more vulnerable to social engineering techniques, which often leverage urgency and fear of authority. Furthermore, the negative association with the volume of past attacks suffered refutes the hypothesis that mere past exposure increases immunity; on the contrary, this finding could indicate an effect of habituation to threats or a pre-existing structural vulnerability on the part of the user.

A central finding from the study concerns the role of time in email management. The negative correlation between response times and self-reported impulsivity provides behavioral validation of the psychological construct under investigation. The marked, statistically significant difference found confirms that victims process information and make decisions much more rapidly than resilient individuals. Slowness and deliberation thus emerge as essential defensive cognitive mechanisms: processing stimuli in a less impulsive manner ensures a greater likelihood of identifying semantic and visual indicators of fraud.

The clustering analysis allowed us to translate these analytical findings into two distinct operational profiles, offering a pragmatic interpretation of the dynamics at play. \textbf{Aware User:} Represents the ideal approach. Supported by high Seniority (F1), a methodical, less impulsive task execution strategy, and characterized by low Creativity (F3), this user adopts a highly cautious approach (longer reaction times). This core combination allows them to achieve the highest phishing identification performance. Additionally, their solid supplementary traits, such as higher technical Expertise (F2), contribute to a strong psychological resilience against common social engineering triggers. \textbf{High-Risk User:} This category encompasses the most critical cognitive-behavioral combination, representing the vast majority of the sample. A structural lack of Seniority (F1), combined with marked Creativity (F3) and higher impulsivity (shorter response times), results in a hasty evaluation process and a lower propensity for critical analysis. Furthermore, their lower baseline technical Expertise (F2) and heightened susceptibility to psychological triggers make this group the primary target and most vulnerable to phishing attacks. This leads us to answer our \textit{RQ}: phishing vulnerability is associated with low seniority (defined as a combination of chronological age and job experience), moderately low Computer Science expertise, and high creativity. We define creativity as a personality dimension that encompasses openness of mind, extroversion, and impulsivity, with the latter being indirectly associated with phishing vulnerability through reduced response times.

From a practical and managerial perspective, the results of this study offer a crucial contribution to the evolution of cybersecurity awareness programs. The results obtained through the K-Means segmentation demonstrates the potential ineffectiveness of standardized (one-size-fits-all) training. Future developments should focus on designing personalized and adaptive training interventions. For instance, programs should be specifically tailored for the High-Risk cluster, incorporating cognitive de-biasing modules aimed at encouraging analytical processing, slowing down decision-making times, and mitigating their inherent psychological vulnerabilities.

Building on these findings, a promising research trajectory involves validating the results in in real-world operational contexts, as well as exploring the geographical factor, as the creation of targeted, country-specific datasets would enable cross-cultural studies, essential for understanding whether and how cultural differences, educational models, or socioeconomic backgrounds influence the perception of cyber risk and the distribution of vulnerability profiles.

\subsubsection{Study Limitations} 
This study has several methodological limitations that should be considered. For example, the use of self-reported measures to assess psychological variables (such as impulsivity and personality traits) inherently exposes the data collected through the questionnaire to potential self-reporting biases or social desirability effects, although this limitation was partially mitigated by the objective measurement of reaction times. Also, administering the recognition task in a simulated environment may not accurately replicate the cognitive load, environmental distractions, or stress associated with daily email management. Regarding data collection, the reliance on a dataset open to the general public provided a large sample but lacked systematic control over geographic origin, preventing the isolation of specific socio-cultural effects. Finally, the cross-sectional design of the study limits the ability to infer definitive causal relationships between the latent constructs and actual effectiveness in phishing recognition. Future research should aim to overcome these limitations, primarily by maximizing ecological validity. 
%Field studies (in the wild) capable of monitoring interaction behaviors with simulated emails within everyday organizational networks would allow for the empirical testing of the two identified profiles (Aware and High-Risk) in real-world operational contexts.
\subsubsection{Ethical Considerations} The analysis presented in this work was conducted on anonymized data. The \textit{Spamley} dataset was collected in the context of a project that received the approval from an ethical committee; the anonymization techniques and more details about the ethical approval of the study can be found in previous work \cite{gallo2024}.

\section{Conclusions}
\label{sectionConclusions}
This study explored the complex interaction between sociodemographic variables, psychological traits, and behavioral patterns in determining susceptibility to phishing attacks. The analyses conducted demonstrated that resilience to cyber threats is not attributable to a single trait, but rather to the synergistic combination of operational maturity (Seniority), methodical evaluation (low Creativity), and technical knowledge (Expertise). At the same time, the behavioral validation confirmed the crucial role of the temporal factor: faster reaction times are systematically associated with a higher error rate, highlighting how deliberation and analytical processing represent indispensable cognitive defenses.

The most significant contribution of the research lies in the empirical profiling of the sample. The segmentation into two distinct operational clusters—Aware User and High-Risk User—clarifies the specific configurations of traits that drive vulnerability. In particular, the characterization of the High-Risk profile—which represents the vast majority of the sample—highlights how a structural lack of operational maturity and technical expertise, compounded by a hasty decision-making approach and high levels of creativity, creates a critical cognitive vulnerability that malicious actors can easily exploit.

Ultimately, the results obtained call for a clear paradigm shift in human risk mitigation strategies within the context of cybersecurity. Organizations and decision-makers are called upon to abandon generic training approaches in favor of targeted and adaptive awareness programs. Promoting true digital security today means not only transferring technical knowledge, but actively addressing cognitive biases, mitigating impulsivity, and training users to engage with the digital space in a more critical and deliberate manner.

% \noindent \textbf{Acknowledgments.} This work was partially supported by project \textit{cyberHuman}, part of the \textit{SERICS} program (PE00000014) under the \textit{MUR National Recovery and Resilience Plan} funded by the \textit{European Union - NextGenerationEU} and by the \textit{European Union - Next Generation EU}, Mission 4, Component 1, through the \textit{ADAPTO} project, part of the \textit{RESTART} program, CUP E63C2 2002040007, CP PE0000001.

% \clearpage
\bibliographystyle{splncs04}
\bibliography{reference}

\end{document}